\documentclass[3p,twocolumn]{elsarticle}

\usepackage{amsmath}
\usepackage{caption}
\usepackage{color}
\usepackage{graphicx}
\usepackage[utf8]{inputenc}
\usepackage{multirow}
\usepackage{subcaption}

\begin{document}

\begin{frontmatter}

\title{Structure of $^{10}$N in $^9$C+p resonance scattering}

\author{J. Hooker}
\author{G.V. Rogachev}
\author{V.Z. Goldberg}
\author{E. Koshchiy}
\author{B.T. Roeder}
\author{H. Jayatissa}
\author{C. Hunt}
\author{C. Magana}
\author{S. Upadhyayula}
\author{E. Uberseder}
\author{A. Saastamoinen}

\address{Department of Physics \& Astronomy and Cyclotron Institute, Texas A\&M University, College Station, TX 77843, USA}

\begin{abstract}
The structure of exotic nucleus $^{10}$N was studied using $^9$C+p resonance scattering. Two $\ell=0$ resonances were found to be the lowest states in $^{10}$N. The ground state of $^{10}$N is unbound with respect to proton decay by 2.2(2) or 1.9(2) MeV depending on the 2$^-$ or 1$^-$ spin-parity assignment, and the first excited state is unbound by 2.8(2) MeV.
\end{abstract}


\end{frontmatter}

There is a permanent interest in studying the most exotic cases of the nuclei chart. New findings  provide a challenge for nuclear structure models under extreme conditions of high isospin and low binding energies.  Among other peculiarities,  exotic nuclei demonstrate unusual shell structures and clustering phenomena, and unusual system sizes which often are  associated with halo configurations. Microscopic {\it ab initio} theoretical approaches can be applied for systems with restricted number of nucleons, therefore light exotic nuclei provide for the best test for a variety of theoretical approaches \cite{Navratil:2009ut}.

$^{10}$N nucleus was a theme of several theoretical works which predicted different spins for the ground state of $^{10}$N (2$^-$, 1$^+$) and instability to proton decay from 1.5 up to 2.8 MeV \cite{Aoyama:1997ymi, Fortune:2013wba, Battye:2009ad, Tilley:2004zz}. However, only one experimental study claimed observation of the $^{10}$N ground state in the $^{10}$B($^{14}$N,$^{14}$B)$^{10}$N reaction \cite{Lepine-Szily:2002gg}. A broad structure at 2.6(4) MeV above the threshold for the $^9$C+p decay was understood as an $\ell$=0 resonance with a width of 2.3(2) MeV. The large width of the state was used as the main argument for $\ell$=0 assignment, the validity of which was questioned in \cite{Tilley:2004zz}.

It is worthwhile to note that many predictions for the $^{10}$N spectrum are based on the  data for the mirror $^{10}$Li nucleus which was studied very extensively (see the latest reviews in \cite{Cavallaro2017, Sanetullaev:2016oiu}). A special interest to the $^{10}$Li spectrum is that the description of  the very neutron rich $^{11}$Li nucleus relies on the nature of the interaction between one neutron and the $^9$Li core.  However, presence of the lowest  s-wave strength which was observed in nearby N=7 nuclei is still questionable in $^{10}$Li \cite{Cavallaro2017}.

We performed an experimental study of the $^{10}$N structure with an idea to also contribute to resolving the puzzle of the mirror $^{10}$Li nucleus.  The nuclear instability of $^{10}$N is well established. So, the lowest $^{10}$N states were populated in resonance $^9$C+p scattering.

The $^9$C beam was produced in the $^{10}$B(p,2n) reaction and separated from the primary beam and other reaction products using the experimental facilities of the Cyclotron Institute at Texas A\&M University, which include a superconducting cyclotron, a cooled hydrogen gas target with a pressure of 3 atm, and a mass separator MARS \cite{Tribble:1989}. We obtained a 23.4$\pm$0.4 MeV/A beam of $^9$C with intensity of 10$^3$ pps using a primary beam of $^{10}$B at 31 MeV/A with intensity of 200 pnA. The $^9$C beam purity was about 40\% with the main contaminant being $^3$He, which is also produced in the gas cell of the primary target.

\begin{figure}
\centering
\includegraphics[scale=0.25]{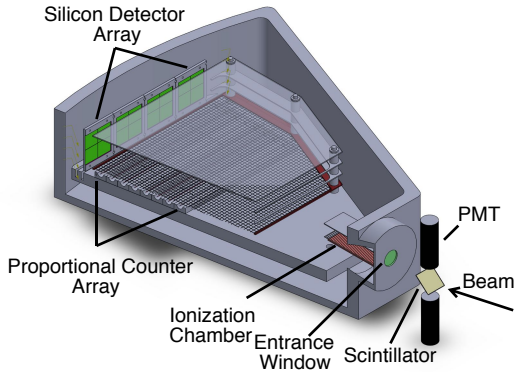}
\caption{(Color online) Sketch of the experimental setup. The BC-404 scintillator is placed in front of the scattering chamber and is observed by the set of two PMTs. The Havar window of 4 $\mu$m thickness separates the gas volume of the scattering chamber and the beamline vacuum. A windowless ionization chamber is installed at the entrance of the scattering chamber. An array of proportional wire cells and a set of silicon detectors are located downstream. \label{fig:chamber}}
\end{figure}

\begin{figure}
\centering
\includegraphics[scale=0.4]{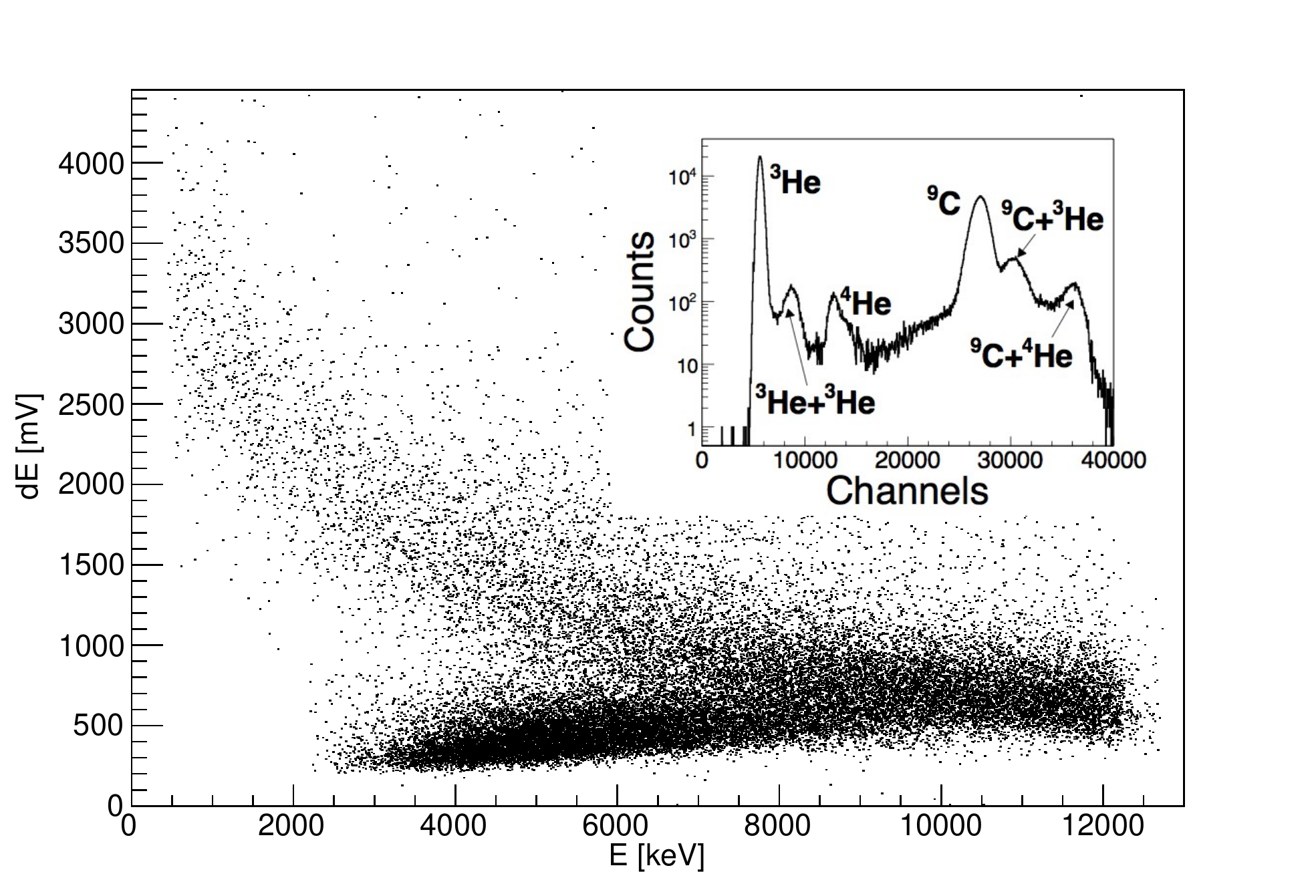}
\caption{The scatter plot of energy deposited in the proportional counter cells (dE) plotted versus the energy measured in the silicon detectors. We divide this into three distinct regions: non-punch-through region, a mixed region where we are uncertain to whether the event is punch-through or not and then a clear punch-through region. The inset shows upstream PMTs energy spectrum obtained  by summing energy signals of the top and bottom PMTs. \label{fig:particleID}}
\end{figure}

The Thick Target Inverse Kinematics (TTIK) method \cite{Artemov:1990, Goldberg:2010zz} was applied to study resonances in $^{10}$N in the $^9$C+p resonance scattering. In the TTIK method, a beam of heavy ions is stopped in the target material, and the light recoil products of the elastic scattering reaction (protons in our case) come out of the interaction region due to the much smaller specific energy loss and are detected. Methane gas (CH$_4$, research grade 99.999\% purity) at pressure of 760 Torr was used for these measurements. This method allows the measurement of a continuous excitation function from the initial energy down to the lowest detectable energy at a single energy of the beam. Because of the method's high efficiency and large resonance cross sections the reasonable statistics can be achieved even at the very low intensities of the rare isotope beams. However, the initial energy of the $^9$C beam of 23 MeV/A is too high to make reliable investigation of states in $^{10}$N which are only a few MeV above the proton decay threshold. The approach similar to the one described in \cite{Goldberg:2010zz} was used to slow down the beam.  BC-404 scintillator of 1 mm thickness was placed upstream of the scattering chamber (Fig. \ref{fig:chamber}). The energy of the beam was degraded down to 9.3 MeV/A in this way. Two  photomultiplier tubes (PMT) were used to collect the light signals from the scintillator. The signals from the PMTs were used to monitor the beam, and to obtain a ``start'' signal for time-of-flight (TOF) measurements, and to identify the main beam contamination that was not filtered out by the velocity filter of MARS. The energy spectrum obtained by summing the signals from the two PMTs is shown in the inset of Fig. \ref{fig:particleID}. It provides clear discrimination between the $^9$C ions and other contaminants (see inset of Fig \ref{fig:particleID}). A windowless ionization chamber (ICE) was placed in the scattering chamber close to the entrance window (see Fig. \ref{fig:chamber}) to measure the specific energy loss of incoming ions and to provide for the final overall normalization of the cross sections. The main improvement of the setup as compared to that of Ref. \cite{Goldberg:2010zz} was construction of the resistive wire -based time projection chamber (TPC) which allowed for 3D tracking of the light recoils and particle ID (see Fig. \ref{fig:particleID}). Only light recoils reached the tracking region, while heavy recoils and $\sim$90\% of $^9$C beam ions were stopped before the active TPC region. Tracking improves energy resolution when the beam is broad in space and energy and helps to reduce background associated with non-elastic scattering events. Three MSQ25-1000 MicronSemiconductor Si detectors were used. Each Si detector consisted of four 25x25 mm$^2$ square segments. Protons deposit a maximum energy of about 12.5 MeV in the silicon detector before they punch-through. The energy of the punch-through protons were reconstructed by identifying the punch-through events using tracking, specific energy loss in the time projection chamber and energy deposited in the Si detector. The final energy spectrum of protons can then be divided into three regions. The first region corresponds to c.m. energies below 3.2 MeV. Recoil protons in this region are stopped in the Si detector. Energy resolution of the proton spectra in this energy region is about 40 keV (here and below GEANT4 Monte Carlo simulation was used to determine c.m. energy resolution). The second region is between 3.2 and 4.5 MeV. This region corresponds to proton energies close to the punch-through energy and is the region that is most sensitive to the assumptions made in the total energy and track reconstruction procedure. While we used the complete Monte Carlo simulation to guide the reconstruction, it is impossible to guarantee accuracy of the measured cross section at all angles in this region. The overall gross features are correct, while the fine details and especially narrow structures (if any are actually present) may be washed out by the reconstruction. This is the least reliable energy region in our data. The third energy region is from 4.5 to 5.5 MeV. Protons that correspond to this energy region can reliably be identified as punch-through events. We estimate that energy resolution in this region is about 200 keV. The excitation function at c.m. energies above 5.5 MeV is affected by the fact that energy distribution of the $^9$C beam as it enters the scattering chamber is rather broad and progressively less $^9$C ions have the energy high enough to populate states in $^{10}$N at 5.5 MeV and above. As a result, the measured cross section trends down after 5.5 MeV. While Monte Carlo simulation was used to make a correction, we decided to exclude the higher energy region from the consideration in this Letter.    

\begin{figure}
  \centering
  \begin{subfigure}[b]{0.6\textwidth}
    \includegraphics[scale=0.4]{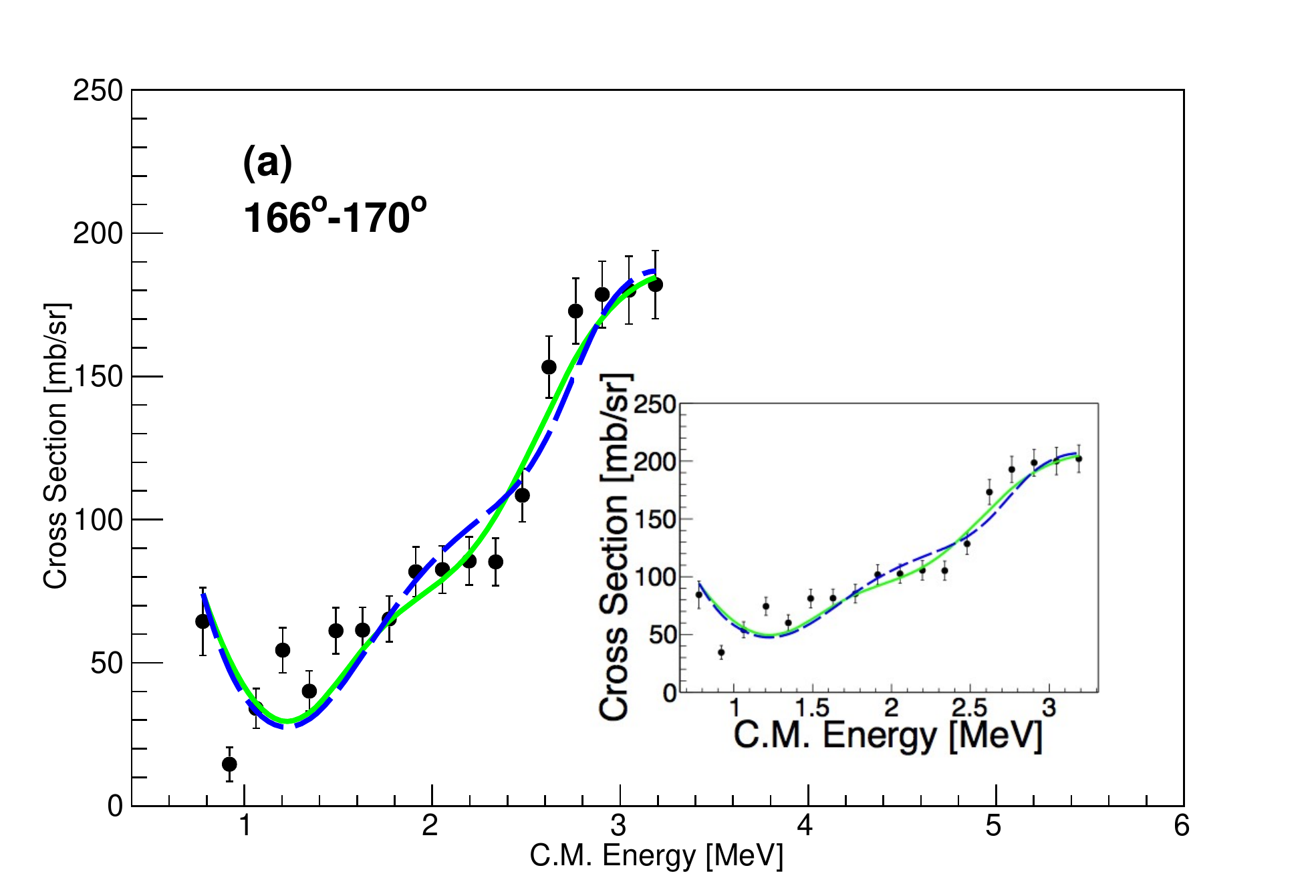}
  \end{subfigure}
  \\
  \begin{subfigure}[b]{0.6\textwidth}
    \includegraphics[scale=0.4]{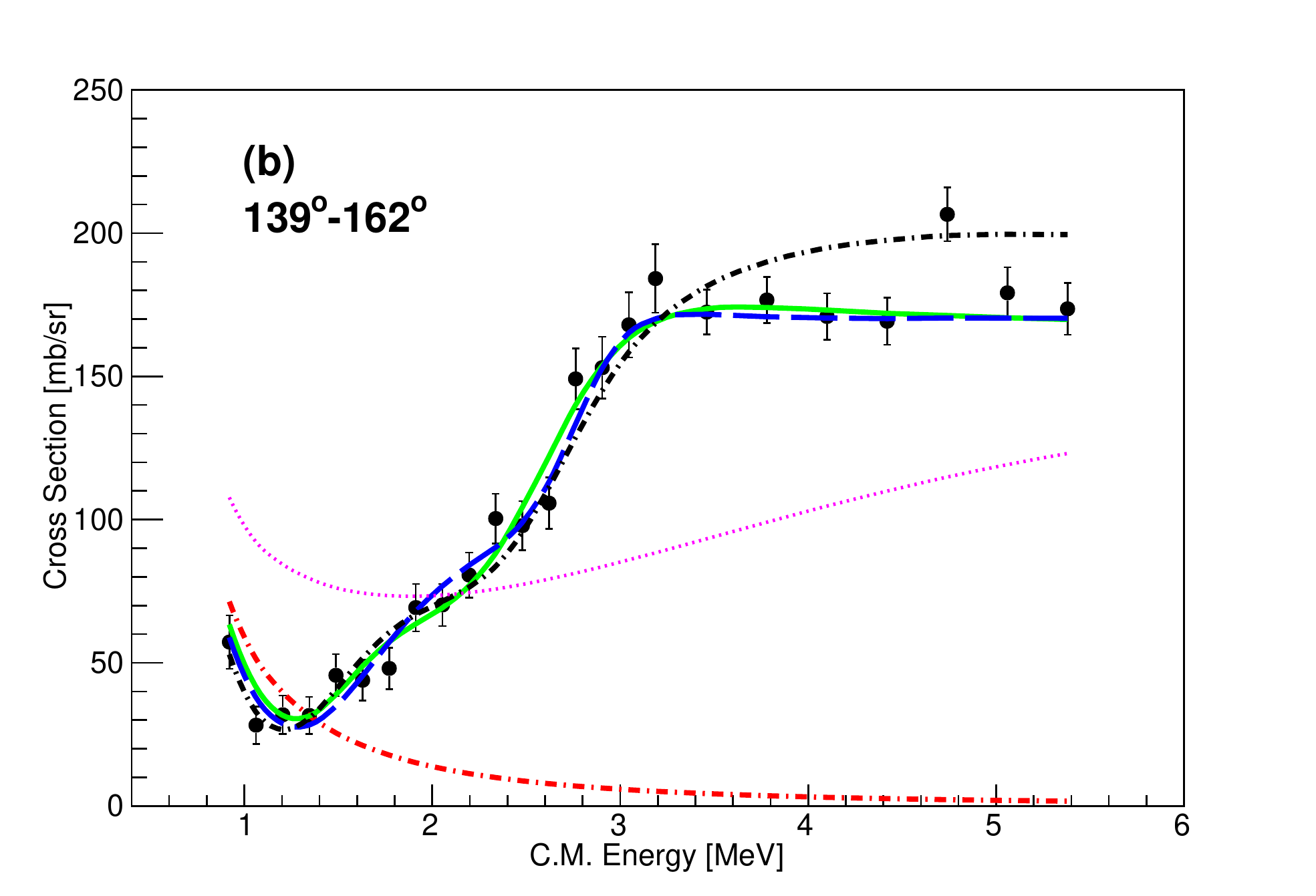}
  \end{subfigure}
  \\
  \begin{subfigure}[b]{0.6\textwidth}
    \includegraphics[scale=0.4]{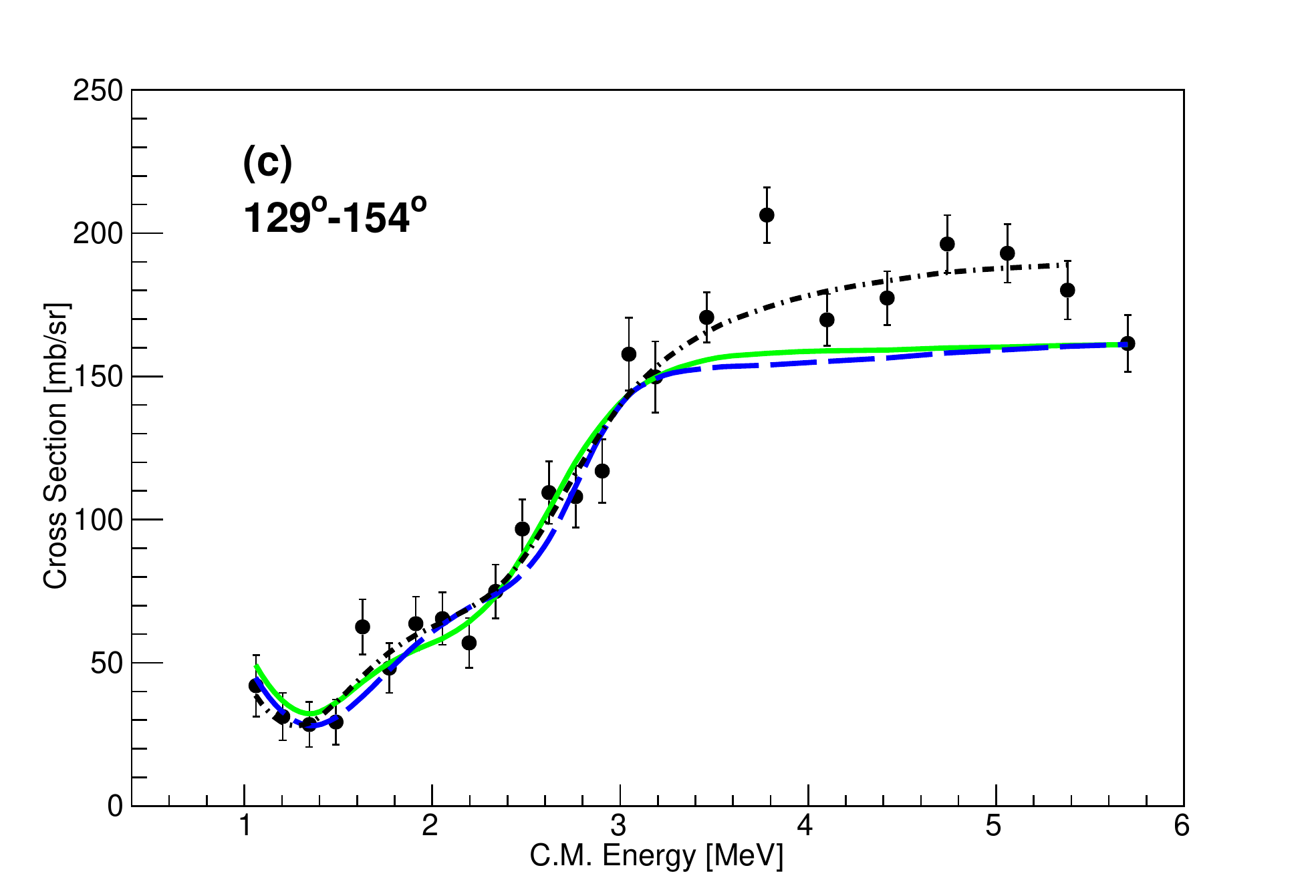}
  \end{subfigure}
  \caption{(Color online) Spectrum of protons from the $^9$C+p resonance scattering at three different angular regions. The red dash-dotted line in (b) is the Rutherford scattering cross section, the blue dashed line is the best fit with 2$^{-}$ as the ground state and a 1$^{-}$ state as the 1st excited state. The green solid line is the best fit with a 1$^{-}$ ground state and a 2$^{-}$ 1st excited state. The dotted magenta curve is the R-matrix calculation assuming that there are no resonances in the $^9$C+p system, and that the cross section is defined by the repulsing hard sphere phase shift. The black dash-dotted curve is the fit with a 2$^-$ g.s., a 1$^-$ 1st excited and a 1$^+$ state at 3.3 MeV (see text for more details). The inset in (a) shows data at the most forward angle without 25 mb/sr background subtraction. \label{fig:spectra}}
\end{figure}

The spectrum of protons from the $^9$C+p resonance scattering is shown in Fig. \ref{fig:spectra} at three scattering angle sets. The most forward angle data is restricted to the first energy region because the track reconstruction procedure does not lead to the reliable reaction vertex localization at the most forward angles and subsequently inhibits punch-through identification. Due to the use of extended gas target the measured scattering angle is a function of energy. The scattering angle values given in Fig. \ref{fig:spectra} correspond to the average scattering angle in c.m. with the smaller values corresponding to the lower c.m. energy. Absolute normalization of the cross section is done by counting every incoming $^9$C ion in the ICE. Statistical uncertainty is dominating the cross section error budget.

\begin{figure}
\centering
\includegraphics[scale=0.4]{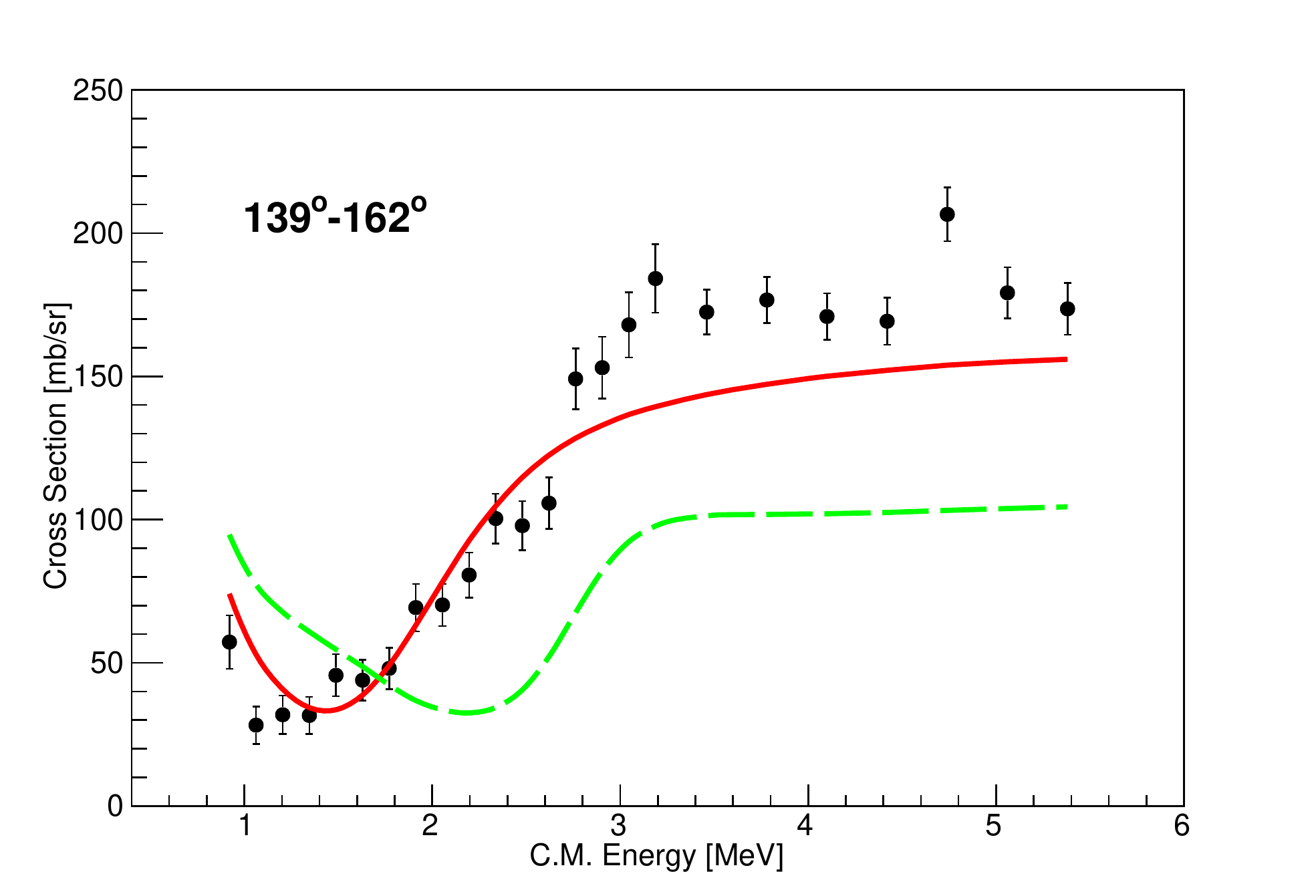}
\caption{(Color online) The $^9$C+p elastic scattering excitation function in the 139-162$^{\circ}$ angular range. The red solid curve is the R-matrix calculation with only one state - the 2$^-$ resonance. The green dashed line is an $\ell=1$ 1$^{+}$ ground state at 2.2 MeV with a 1$^{-}$ state as the 1st excited state. \label{fig:fits}}
\end{figure}

Two observations can be made without a detailed analysis. The excitation function is mostly featureless except for the low energy rise of the cross section from 1 to 3 MeV in c.m. and it is much higher than the pure Rutherford scattering shown by the red solid curve in Fig. \ref{fig:spectra} b) at all energies above 1.5 MeV. These properties of the observed spectra indicate an important role of the $\ell=0$ resonances. Only $\ell=0$ resonances can provide for broad structures at low energies. The multi-channel multi-level R-matrix analysis was performed using code MinRMatrix \cite{JohnsonThesis}. In order to minimize the number of free parameters we only used $\ell=0$ and $\ell=1$ partial waves. This is justified by low energy and light mass of the colliding nuclei. If no resonances present in the measured energy range then the cross section is defined by the repulsive hard sphere phase shift that in turn is determined by a single parameter - channel radius. This cross section that has no resonances is shown in Fig. \ref{fig:spectra}(b) by dotted magenta curve, and it does not reproduce the experimental data. In order to describe the rapid rise of the cross section near 2 MeV, we introduced a 2$^-$ $\ell=0$ state of single particle nature in $^{10}$N at 2.2 MeV. While this resonance provides for the needed rise, this is insufficient to account for the measured cross section, as shown with the red solid curve in Fig. \ref{fig:fits}. The s-wave is split between two spin-parities, the 2$^-$ and the 1$^-$. We introduced the second $\ell=0$ state, the 1$^-$ at 2.8 MeV, to account for the high experimental cross section and to provide for the correct shape of the excitation function. The corresponding curve is shown as a dashed blue line in Fig. \ref{fig:spectra}. We were also able to fit the excitation function using a 1$^-$ $\ell=0$ state at 1.9 MeV and a second $\ell=0$ state, the 2$^-$ at 2.8 MeV. This is the solid green curve in Fig. \ref{fig:spectra}. There are only 4 free parameters in the R-matrix fits shown in Fig. \ref{fig:spectra}: the energy and the reduced width for the 1$^-$ and 2$^-$ states. We used 5.0 fm as a channel radius parameter. Of course, it is possible to fit the data with any other reasonable choice of channel radius by introducing ``background'' resonances at higher excitation energies and the $\ell=2$ partial waves. We verified that it has negligible effect on the parameters of the two $\ell=0$ states. An attempt to fit the excitation function using the positive parity assignment for the ground state fails due to characteristic interference pattern between the p-wave and the Coulomb amplitude (see the green dashed curve in Fig. \ref{fig:fits}). The parameters of the R-matrix best fit are given in Table \ref{tab:table1}. It was found that small non-coherent background at the level of 25 mb/sr is necessary to achieve a good fit. The likely sources of this background are fusion-evaporation reaction $^9$C+$^{12}$C in the gas and breakup of $^9$C on nuclei of gas atoms and entrance window material. Similar background was measured previously in \cite{Goldberg:2010zz} and was determined that it is almost uniform and effectively contributed up to 30-40 mb/sr. We used the results of Ref. \cite{Goldberg:2010zz} as a guide and assumed that the background observed in similar experimental conditions (with the rare isotope beam of $^{13}$O which is similarly neutron deficient) is the same. Also, due to favorable experimental conditions at 139-162$^{\circ}$ scattering angles and with the help of tracking we were able to do cleaner cuts and reduced the contribution of this background to negligible level in this angular range. Therefore, no background was subtracted from the experimental data in Fig. \ref{fig:spectra} (b), and 25 mb/sr background was subtracted from the experimental data shown in Fig.  \ref{fig:spectra} (a,c).

The following main conclusions can be made from the R-matrix analysis described above. The low energy spectrum of protons from the $^9$C+p resonance scattering is defined by the s-wave states. It is not possible to fit the shape and magnitude of the excitation function without introducing both of the $\ell=0$ resonances. We have not been able to make a definitive spin assignment for these 2s states. If the spin-parity of the ground state is 2$^-$ then its best fit c.m. energy is 2.2 MeV, and 1.9 MeV in case of 1$^-$ spin-parity assignment. In either case the g.s. has single-particle nature with dimensionless reduced width around 0.8 (see Table \ref{tab:table1}). The first excited state, however, has significant contributions from channels other than $^9$C(g.s.)+p (more so if the 2$^-$ g.s. and the 1$^-$ 1st excited is the correct sequence). This fact can be used to make a definitive spin-parity assignment if coupled to rigorous theoretical analysis of the $^{10}$N/$^{10}$Li system.

There is no clear evidence for an $\ell=1$ state(s) in the measured spectrum.  Based on the latest experimental information, the lowest p-wave state is observed at 0.45(3) MeV above the neutron decay threshold in $^{10}$Li and has width of 680(30) keV \cite{Cavallaro2017}. The potential model calculations (details are provided below) give the mirror state in $^{10}$N at around 3.3 MeV above the proton threshold. The quality of our data in this energy region is handicapped significantly due to the punch-through problem. We included a p-state, the 1$^+$, at 3.3 MeV as an illustrative example in Fig. \ref{fig:spectra} (c) (the black dash-dotted curve). The reduced width for this 1$^+$ state was adjusted to reproduce its width in $^{10}$Li \cite{Cavallaro2017}. We verified that such state at 3.3 MeV (or above) does not contradict the measured excitation function. However, we do not claim observation of any p-wave states due to uncertainties in c.m. energy reconstruction, limited energy resolution in the energy region above 3.2 MeV, and uncertainties related to contributions of other higher lying states and $\ell>1$ partial waves into the cross section above 3 MeV.

Our data should be compared with the results of the only work that claimed observation of the $^{10}$N ground state at 2.6 MeV \cite{Lepine-Szily:2002gg} in the $^{10}$B($^{14}$N,$^{14}$B)$^{10}$N reaction. This energy is between the g.s. and the first excited state of $^{10}$N observed in this work. It is possible that one (or both) of these states was in fact observed in Ref. \cite{Lepine-Szily:2002gg}. The authors of Ref. \cite{Tilley:2004zz} questioned this assignment on the basis of the systematics of the mass transfer reactions, that predominantly populate $\ell$=1 states instead of a broad $\ell$=0 resonances. Of course, measurements with better counting statistics and an analysis which takes into account two overlapping s-states are highly desirable to verify if these structures can really be populated in the mass transfer reactions like $^{10}$B($^{14}$N,$^{14}$B)$^{10}$N.  

{\it Structure of $^{10}$Li and $^{9}$N.} Below we make extrapolations that are motivated by the results of this work for the structure of $^{10}$Li, and $^9$N isotopes, focusing on the location of the 2s1/2 shell. Since both of the low lying negative parity states in $^{10}$N have been observed, the location of the 2s-shell in $^{10}$N can be constrained rather well. Taking into account uncertainty related to the spin-parity assignment the 2s shell is unbound by 2.3$\pm$0.2 MeV in $^{10}$N. One can use potential model to evaluate the location of the 2s shell in the mirror nucleus - $^{10}$Li. Using conventional parameters of the Woods-Saxon potential (r=1.25 fm, a=0.7 fm, and r$_c$=1.3 fm) that describe the $^{11}$N/$^{11}$Be pair reasonably well, we can conclude that the 2s shell is located near 100 keV above the neutron decay threshold in $^{10}$Li and due to the energy splitting between the 2$^-$ and 1$^-$ the ground state of $^{10}$Li should then be the 2s state located between 0 and 100 keV (probably within few tens of keV above the neutron threshold). Therefore, while in the recent $^9$Li($^2$H,p) experiment \cite{Cavallaro2017} the $\ell=0$ states close to the neutron threshold in $^{10}$Li were not found, our results on $^{10}$N support the existence of the lowest $\ell=0$ states in $^{10}$Li. Moving further beyond the nuclear stability, the location of the 2s1/2 shell in $^9$N can be established using linear extrapolation similar to that suggested in Ref. \cite{Talmi:1960zz}. Specifically, the 2s1/2 shell is located at 1.49(6) MeV in $^{11}$N above the proton threshold according to the recent compilation \cite{Kelley2012}. (Although it varies from 1.3 MeV to 1.6 MeV in the literature, see references in \cite{Kelley2012}). Removing one neutron ``moves'' the proton 2s1/2 shell to 2.3 MeV in $^{10}$N and therefore to about 3.0-3.3 MeV in $^{9}$N, with the two neutrons removed from the $^{11}$N. This implies that $^9$N is unbound with respect to six particle decay (5 protons and $\alpha$-particle) by 6.5 MeV or more, making it unlikely that the 2s ``ground state'' of $^9$N will ever be conclusively observed.

The main result of this Letter is that the ground and the first excited states in $^{10}$N have been investigated in detail for the first time. While definitive spin-parity assignments for these two states cannot be concluded from this data, it can be shown that these states are definitely s-wave resonances. $^{10}$N is unbound by 1.9$\pm$0.2 MeV assuming a 1$^-$ spin-parity assignment for the ground state or by 2.2$\pm$0.2 MeV if the ground state is 2$^-$. The first excited state at 2.8 MeV is also an s-wave (1$^-$ or 2$^-$, depending on the ground state spin parity assignment). The ground state is a single particle state, whereas the first excited state has significant contribution from configurations other than the $^9$C(g.s.)+p. No evident p-wave resonances were observed. However, it was demonstrated that their existence at or above 3.3 MeV does not contradict the experimental data. Observation of the two s-wave states in $^{10}$N constrains the location of the 2s1/2 shell in $^{10}$N. This information can be used to make extrapolations for the location of the 2s1/2 shell in more exotic $^9$N isotope and also in the mirror $^{10}$Li nucleus. The $^9$N appears to be unbound by about 3 MeV with respect to the proton emission and by about 6.5 MeV with respect to 6 particle decay ($\alpha$+5p). The ground state of $^{10}$Li has to be the s-wave resonance at energy very close to the neutron decay threshold (definitely within 100 keV and most likely within few tens of keV from the threshold).


\renewcommand{\arraystretch}{2.}
\begin{table*}
\centering
\begin{tabular}{|c | c | c | c | c|}
\hline
state & J$^{\pi}$ & E$^*$ [MeV] & $\Gamma$ [MeV] & $\Theta^{2}$ \\ \hline
\multirow{2}{*}{G.S.} & 2$^{-}$ & $2.2\substack{+0.2 \\ -0.2}$ & $3.1\substack{+0.9 \\ -0.7}$ & $0.8\substack{+0.2 \\ -0.2}$ \\ \cline{2-5}
  & 1$^{-}$ & $1.9\substack{+0.2 \\ -0.2}$ & $2.5\substack{+2.0 \\ -1.5}$ & $0.8\substack{+0.5 \\ -0.3}$ \\ \hline
\multirow{2}{*}{1st excited} & 1$^{-}$ & $2.8\substack{+0.2 \\ -0.2}$ & $1.2\substack{+0.6 \\ -0.4}$ & $0.25\substack{+0.1 \\ -0.1}$ \\ \cline{2-5}
  & 2$^{-}$ & $2.8\substack{+0.2 \\ -0.2}$ & $2.0\substack{+0.7 \\ -0.5}$ & $0.4\substack{+0.2 \\ -0.1}$ \\ \hline
\end{tabular}
\\
$^*$ The phase shift never reaches 90$^{\circ}$ for these broad $\ell=0$ states. Resonance energy is defined as the ``peak'' energy, that is energy at which the cross section is maximum if Coulomb amplitude and influence of other resonances is removed.
\caption{Parameters of the R-Matrix best fit for the two lowest state in $^{10}$N. For each configuration, the spin-parity assignment, the energy with respect to the proton decay threshold, the total width of the state, and the dimensionless reduced width are shown. The latter is calculated as ratio between the proton reduced width and single particle reduced width calculated at channel radius of 5.0 fm. \label{tab:table1}}
\end{table*}

The authors are grateful to the cyclotron team at the Cyclotron Institute for consistently reliable operation. The authors acknowledge that this material is based upon their work supported by the U.S. Department of Energy, Office of Science, Office of Nuclear Science, under Award No. DE-FG02-93ER40773. The authors G.V.R. and H.J. are also supported by the Welch Foundation (Grant No. A-1853).

\bibliographystyle{elsarticle-num}
\bibliography{mybib}{}

\end{document}